\documentclass{article}

\usepackage{microtype}
\usepackage{graphicx}
\usepackage{subfigure}
\usepackage{booktabs}
\usepackage{multirow}

\usepackage{hyperref}

\usepackage{algorithmic}

\usepackage[accepted]{mlsys2025}

\usepackage{amsmath}
\usepackage[T1]{fontenc}
\usepackage{microtype}
\usepackage{enumitem}
\setlist[itemize]{leftmargin=*}
\setlist{nosep}

\usepackage[skip=0.15in]{caption}
\usepackage{stfloats}

\newcommand{\sysname}{Arrow}

\newcommand{\parabf}[1]{\noindent\textbf{#1}}

\newcommand{\rom}[1]{\uppercase\expandafter{\romannumeral#1}}
\newcommand{\RETURN}{\STATE {\bfseries return} }

\mlsystitlerunning{Arrow: Adaptive Scheduling Mechanisms for Disaggregated LLM Inference Architecture}

\begin{document}

\twocolumn[
\mlsystitle{Arrow: Adaptive Scheduling Mechanisms for Disaggregated LLM Inference Architecture}

\begin{mlsysauthorlist}
\mlsysauthor{Yu Wu}{ustc}
\mlsysauthor{Tongxuan Liu}{ustc}
\mlsysauthor{Yuting Zeng}{ustc}
\mlsysauthor{Siyu Wu}{buaa}
\mlsysauthor{Jun Xiong}{jd}
\mlsysauthor{Xianzhe Dong}{ustc}
\mlsysauthor{Hailong Yang}{buaa}
\mlsysauthor{Ke Zhang}{jd}
\mlsysauthor{Jing Li}{ustc}
\end{mlsysauthorlist}

\mlsysaffiliation{ustc}{University of Science and Technology of China}
\mlsysaffiliation{buaa}{Beihang University}
\mlsysaffiliation{jd}{JD.com}

\mlsyskeywords{Inference Serving; Prefill-Decode Disaggregation; Request Scheduling}

\vskip 0.3in

\begin{abstract}
Existing large language model (LLM) serving systems typically employ Prefill-Decode disaggregated architecture to prevent computational interference between the prefill and decode phases. However, in real-world LLM serving scenarios, significant fluctuations in request input/output lengths lead to imbalanced computational loads between prefill and decode nodes under traditional static node allocation strategies, consequently preventing efficient utilization of computing resources to improve the system's goodput. To address this challenge, we design and implement \sysname{}, an adaptive scheduler that leverages stateless instances and latency characteristics of prefill and decode tasks to achieve efficient adaptive request and instance scheduling. \sysname{} dynamically adjusts the number of instances handling prefill and decode tasks based on real-time cluster performance metrics, substantially enhancing the system's capability to handle traffic spikes and load variations. Our evaluation under diverse real-world workloads shows that \sysname{} achieves up to $2.55 \times$ higher request serving rates compared to state-of-the-art Prefill-Decode disaggregated serving systems.
\end{abstract}
]

\printAffiliationsAndNotice{}

\section{Introduction}
\label{sec:introduction}

Large language models (LLMs), such as Gemini~\cite{teamGeminiFamilyHighly2024,teamGemini15Unlocking2024}, GPT~\cite{openaiGPT4TechnicalReport2024,openaiGPT4oSystemCard2024}, Llama~\cite{touvronLlama2Open2023,grattafioriLlama3Herd2024}, Qwen~\cite{baiQwenTechnicalReport2023,qwenQwen25TechnicalReport2025}, and DeepSeek~\cite{deepseek-aiDeepSeekR1IncentivizingReasoning2025,deepseek-aiDeepSeekV3TechnicalReport2025}, have achieved remarkable success across various domains. Building upon these models, a series of innovative applications have emerged, including LLM-based search engines~\cite{perplexityinc.Perplexity2025,mehdiReinventingSearchNew2023}, personal assistants~\cite{openaiinc.ChatGPT2025,deepseekinc.DeepSeek2025}, and embodied intelligence systems~\cite{wangVoyagerOpenEndedEmbodied2023,liCAMELCommunicativeAgents2023}, highlighting the tremendous application potential of LLMs. However, deploying these LLMs’ inference services in production environments presents numerous challenges. The massive parameters of LLMs and ultra-long input sequences impose tremendous computational and memory demands, making it difficult to consistently meet strict Service Level Objectives (SLOs) even when using high-performance GPU servers~\cite{zhouSurveyEfficientInference2024,yuanLLMInferenceUnveiled2024,zhenTamingTitansSurvey2025}.

In LLM inference services, token generation proceeds iteratively in an autoregressive manner, where each iteration decodes the next token based on previous token sequence. The inference process is typically divided into two distinct phases: prefill and decode. During the prefill phase, the entire input token sequence undergoes forward propagation to generate the first output token. In the decode phase, each newly generated token is concatenated to the end of the current token sequence and used as input for generating the subsequent token in each iteration. Two key metrics are commonly used to evaluate the performance of inference services: (1) Time-to-First-Token (TTFT), measuring the latency to generate the first token in the prefill phase; and (2) Time-per-Output-Token (TPOT), representing the average token generation latency in the decode phase. Efficient inference services must satisfy strict SLOs under limited hardware resources while maximizing the system’s goodput.

\begin{figure*}[t]
    \centering
    \includegraphics[width=\linewidth]{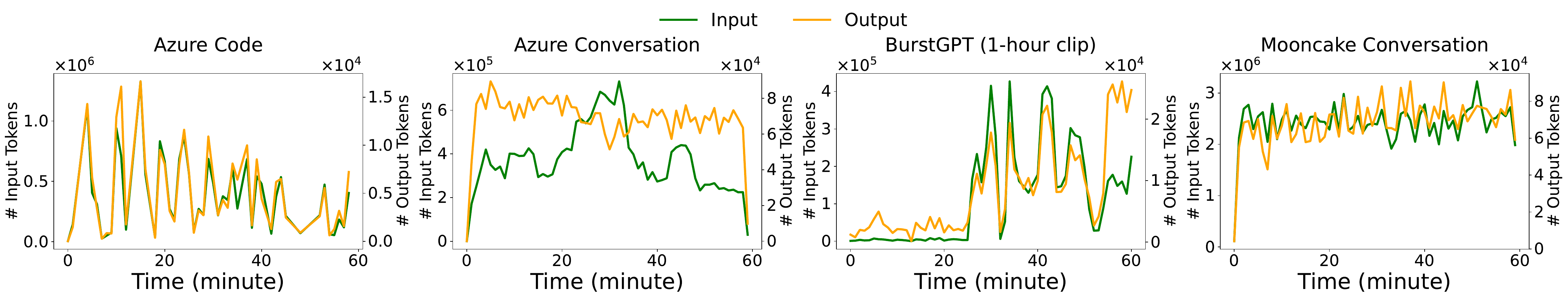}
    \caption{Total request input/output length per minute over time in different LLM serving systems.}
    \label{fig:tokens-over-time}
\end{figure*}

Recently, researchers have proposed various optimization techniques to improve the overall throughput of serving systems and meet SLOs of diverse applications. A notable advancement is continuous batching~\cite{yuOrcaDistributedServing2022}, which implements iteration-level scheduling to dynamically add or remove requests in the batch during each computation iteration. This approach provides greater flexibility compared to traditional request-level scheduling, significantly reducing queuing latency. Subsequent studies introduced chunked prefill~\cite{holmesDeepSpeedFastGenHighthroughputText2024, agrawalTamingThroughputLatencyTradeoff2024}, which divides input sequences into smaller chunks and batches them with decode tasks to mitigate the latency spikes caused by lengthy prompts, further enabling stall-free scheduling that allows new requests to be added without interrupting ongoing decode processing. However, recent works~\cite{zhongDistServeDisaggregatingPrefill2024,patelSplitwiseEfficientGenerative2024,huShuffleInferDisaggregateLLM2025} have shown that prefill and decode phases exhibit fundamentally distinct computational characteristics and latency requirements. Colocating prefill and decode computation from different requests creates mutual interference, causing increased TTFT and TPOT that ultimately degrades the system's goodput.

To address the interference between prefill and decode phases, DistServe~\cite{zhongDistServeDisaggregatingPrefill2024} assigns these phases to separate instances, eliminating phase interference while enabling independent optimization of parallelization strategies for each phase. Splitwise~\cite{patelSplitwiseEfficientGenerative2024} further explores both homogeneous and heterogeneous cluster deployments to optimize cost-efficiency and throughput. ShuffleInfer~\cite{huShuffleInferDisaggregateLLM2025} designs a two-level scheduling algorithm based on resource utilization prediction to prevent decode scheduling hotspots. While disaggregation resolves phase interference, a key challenge remains: properly configuring the ratio between prefill and decode workers to maximize goodput. Improper configuration ratio can lead to severe performance degradation~\cite{qinMooncakeTradingMore2025}.

We observe substantial variability in input and output lengths across real-world LLM inference workloads. This observation is drawn from diverse production traces as illustrated in Figure~\ref{fig:tokens-over-time}, including Azure LLM inference services~\cite{patelSplitwiseEfficientGenerative2024}, Azure OpenAI GPT service (BurstGPT)~\cite{wangBurstGPTRealWorldWorkload2025}, and Kimi conversation service (Mooncake Conversation)~\cite{qinMooncakeTradingMore2025}. The variation of input/output length directly impacts the workload distribution between prefill and decode nodes~\cite{zhongDistServeDisaggregatingPrefill2024, duEcoServeEnablingCosteffective2025}, making the optimal prefill-decode (PD) ratio configuration highly sensitive to workload patterns. Consequently, static PD ratio fails to achieve optimal performance under such fluctuating conditions, necessitating adaptive resource allocation strategies.

To address the above challenge, we first conduct an in-depth analysis of workload variations in real-world inference services. Our study reveals that existing Prefill-Decode disaggregated systems exhibit lagging instance scheduling when handling dynamic workload changes (\S\ref{sec:motivation}). Based on the request processing workflow of Prefill-Decode disaggregated systems, we derive crucial insights for request and instance scheduling (\S\ref{sec:analysis}). Building upon these analyses and insights, we design \textbf{\sysname{}}, an adaptive request and instance scheduler that dynamically schedules requests and instances based on SLO settings and instance load (\S\ref{sec:design}). \sysname{} employs stateless instances where each instance can process both prefill and decode requests without dedicated roles. 
The system features an SLO-aware scheduling algorithm where a global scheduler dynamically adjusts request dispatching and instance allocation based on: (1) predicted TTFT for incoming requests, (2) real-time token generation intervals of ongoing requests, and (3) target SLO metrics.

We implement \sysname{} based on vLLM~\cite{kwonEfficientMemoryManagement2023} and evaluate its performance using diverse production workloads (\S\ref{sec:evaluation}). Experimental results show that \sysname{} can significantly outperform existing approaches, delivering $1.59 \times$\textasciitilde$2.55 \times$ higher request serving rates than state-of-the-art PD-disaggregated systems under given SLO constraints.

In summary, our main contributions are as follows:
\begin{itemize}
    \item Identify that fluctuations in input/output lengths can lead to suboptimal goodput under traditional static PD configuration ratio and propose several key insights for more effective request and instance scheduling. 
    \item Design a novel scheduler \sysname{} that enables adaptive request and instance scheduling through stateless instances and latency characteristics of prefill and decode tasks.
    \item Conduct a comprehensive performance evaluation of \sysname{} using real-world workloads, demonstrating the effectiveness of its adaptive scheduling strategy.
\end{itemize}

\begin{figure*}[t]
    \centering
    \includegraphics[width=\linewidth]{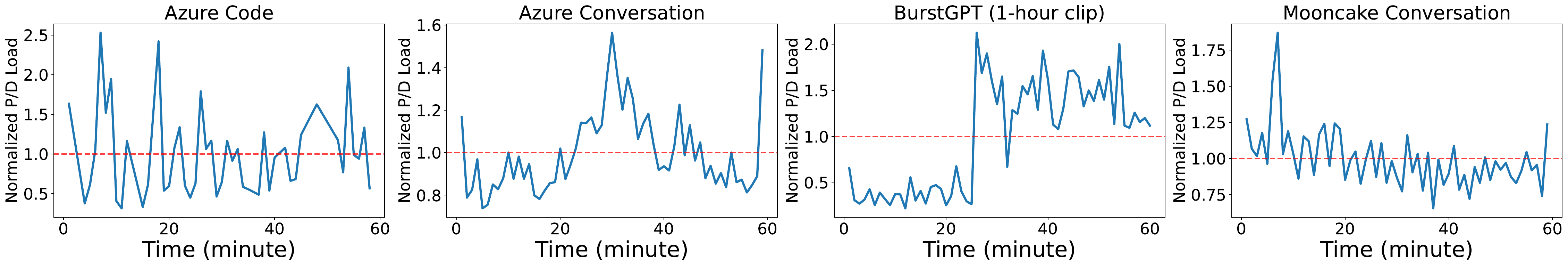}
    \caption{Normalized Prefill / Decode load over time in different traces.}
    \label{fig:load-over-time}
\end{figure*}

\section{Background}
\label{sec:background}

\subsection{LLM Inference}
\label{subsec:background-llm-inference}

Modern LLMs~\cite{openaiGPT4TechnicalReport2024, grattafioriLlama3Herd2024, qwenQwen25TechnicalReport2025, deepseek-aiDeepSeekV3TechnicalReport2025} mostly adopt the transformer architecture~\cite{vaswaniAttentionAllYou2017} and process input sequences through an autoregressive generation process. To avoid redundant computation, existing inference engines typically employ KV Cache~\cite{huggingfaceKVCacheStrategies2025} to cache intermediate results, thereby dividing the computation for a single request into two phases: Prefill and Decode. During the prefill phase, the inference engine processes the user's input, generates KV Cache for all input tokens, and produces the first output token. In the decode phase, the engine computes KV Cache for each newly generated token in subsequent iterations and outputs the next token. Since both prefill and decode phases require shared access to KV Cache, existing serving systems typically colocate these two phases within the same instance. Techniques such as continuous batching~\cite{yuOrcaDistributedServing2022} and chunked prefill~\cite{agrawalTamingThroughputLatencyTradeoff2024} are employed to further optimize the system's overall throughput.

\subsection{Prefill-Decode Disaggregation}
\label{subsec:background-pd-disaggregation}
Recent studies have highlighted significant differences in computational characteristics between the prefill and decode phases~\cite{patelSplitwiseEfficientGenerative2024,zhongDistServeDisaggregatingPrefill2024}, which can lead to mutual interference between these two phases~\cite{huShuffleInferDisaggregateLLM2025} and suboptimal hardware resource allocation~\cite{zhongDistServeDisaggregatingPrefill2024}. To address these issues, researchers have proposed the Prefill-Decode disaggregated inference architecture~\cite{patelSplitwiseEfficientGenerative2024, huShuffleInferDisaggregateLLM2025, zhongDistServeDisaggregatingPrefill2024}, which divides compute instances into two types: prefill instances and decode instances, each dedicated to handling their respective phases. After completing the prefill computation for a request, the prefill instance transfers both the request and its corresponding KV Cache to a decode instance via high-speed interconnects such as NVLink or InfiniBand. The decode instance then proceeds with the subsequent decode computations. This architecture eliminates computational interference between the two phases and enables independent optimization of parallelization strategies and resource allocation for each phase by decoupling prefill and decode, further improving the flexibility in performance tuning and overall goodput.
\section{Motivation}
\label{sec:motivation}
The Prefill-Decode disaggregated inference architecture enables independent optimization for both phases. However, we observe that existing disaggregated systems employing static instance partitioning schemes suffer from low hardware resource utilization and inadequate responsiveness to traffic bursts. In this section, we will elaborate on this issue using production traces from real-world LLM serving systems and present our key insights for addressing it.

\subsection{Diversity of Workloads}
\label{subsec:motivation-diversity}
We conduct an in-depth analysis of the four traces mentioned in Section \ref{sec:introduction}. Figure~\ref{fig:tokens-over-time} shows the total input and output lengths of requests per minute over time. We observe that these traces exhibit substantial temporal variations in request input and output lengths within traces, with per-minute lengths differing by more than 50 times between the lowest and highest load periods. Besides, significant variation in workload characteristics can also be observed across traces: Azure Code and BurstGPT exhibit frequent bursts, while Mooncake Conversation maintains relatively stable loads. Load predictability also varies significantly, with Azure Code showing strong input-output length correlation compared to Azure Conversation's weaker correlation.

Prior works~\cite{zhongDistServeDisaggregatingPrefill2024, duEcoServeEnablingCosteffective2025} reveal that the load of prefill phase scales quadratically with the input length, while the load of decode phase grows linearly with the total number of tokens in the batch. This fundamental difference in scaling characteristics leads to distinct load growth rates between prefill and decode instances. The observed diversity in input/output lengths further exacerbates load fluctuations. To illustrate this fluctuation, we aggregate the total prefill and decode processing times of all requests within each minute in Figure~\ref{fig:tokens-over-time}, treating them respectively as the system’s prefill and decode loads. Using the average Prefill/Decode load ratio across the entire trace as a baseline, we compute the per-minute Prefill/Decode load ratio relative to this baseline, as shown in Figure~\ref{fig:load-over-time}. The red line at 1.0 represents the baseline Prefill/Decode load. A value above 1.0 indicates that a higher P/D instance ratio can achieve better serving performance. Conversely, a value below 1.0 implies that a lower P/D ratio is preferred. We can observe significant variation in prefill and decode load dynamics over time, indicating that serving systems must be capable of dynamically adjusting the ratio of prefill and decode nodes to accommodate varying workload patterns. 

\subsection{Existing Solutions}
\parabf{Workload Profiling and Simulation.} Existing works~\cite{jinDServeServingDisaggregated2024,patelSplitwiseEfficientGenerative2024,qinMooncakeTradingMore2025} typically set the PD ratio based on profiling or simulator data. However, profiling-based methods are only effective when request arrival patterns and length distributions remain relatively stable. In situations with substantial load variations, if adequate instances are provisioned based on the peak load of both types of instances, it may lead to idle hardware resources when the load of one type of task is low. Alternatively, if the PD ratio is set according to the average load, the preconfigured ratio may deviate from the actual load during fluctuations, potentially resulting in SLO violations. 

\parabf{Length and Utilization Monitoring.} Another common approach is to dynamically adjust the types of instances based on the request length distribution or instance utilization. However, the instance flipping strategies adopted by current systems~\cite{zhongDistServeDisaggregatingPrefill2024,patelSplitwiseEfficientGenerative2024,huShuffleInferDisaggregateLLM2025} generally suffer from long response times. These approaches typically involve multiple steps, including observation, waiting for flipping conditions, and restarting instances. The entire process often takes several minutes to finish, exhibiting significant scheduling latency. As a result, the serving system struggles to promptly adapt to workload fluctuations. Moreover, this approach introduces additional instance downtime, which further degrades the system's overall serving capacity.

\section{Analysis}
\label{sec:analysis}
In this section, we present several key insights for request and instance scheduling in PD-disaggregated systems. 

\subsection{Request Processing}
\label{subsec:analysis-process}
\begin{figure}[t]
    \centering
    \includegraphics[width=\linewidth]{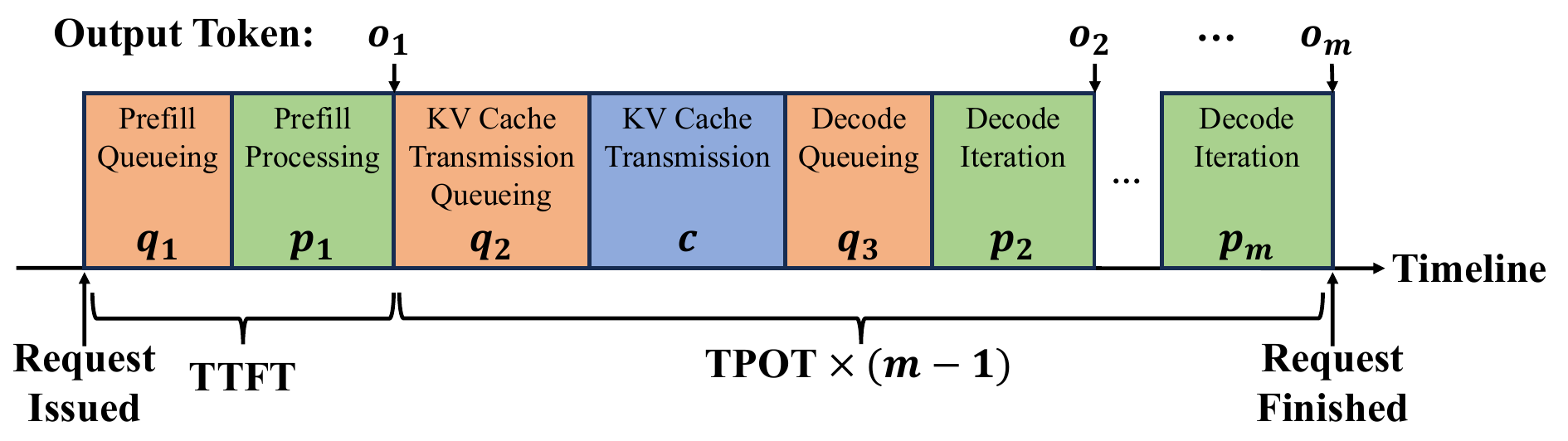}
    \caption{Request processing timeline in Prefill-Decode disaggregated inference system.}
    \label{fig:analysis-timeline}
\end{figure}

Figure~\ref{fig:analysis-timeline} illustrates the complete request processing workflow in a typical PD-disaggregated system. Consider a request $r$ with output tokens $o_1 o_2 \cdots o_m$, the request is first dispatched to a prefill instance, experiencing prefill queuing delay $q_1$ before starting prefill computation (duration $p_1$). The system then waits ($q_2$) for a decode instance to fetch both the request and its corresponding KV Cache, with transfer time $c$. After decode queuing delay $q_3$, the decode instance begins iterative token generation, where each iteration produces one token with computation time $p_2$ through $p_m$. We assume that prefill instances process requests sequentially, while decode instances maximize batch size by grouping multiple decode requests within given batch size and GPU memory constraints. The rationale is that increasing the prefill batch size can hardly bring improvements in throughput; while enlarging the decode batch size can substantially enhance throughput~\cite{zhongDistServeDisaggregatingPrefill2024, patelSplitwiseEfficientGenerative2024, huShuffleInferDisaggregateLLM2025}.

\subsection{TTFT}
\label{subsec:analyais-ttft}
TTFT (Time-to-First-Token) is a key indicator of the processing capability of prefill instances. It is defined as the time from when the user issues a request until the first token is received, corresponding to the $q_1 + p_1$ duration illustrated in Figure~\ref{fig:analysis-timeline}. Suppose there are $n$ prefill requests $r_1 \cdots r_n$ to be processed on the prefill instance. Let $a_i$ denote the arrival time of the $i$-th request, $e_i$ denote its computation completion time, $q_1^{(i)}$ denote the prefill queuing delay, and $p_1^{(i)}$ denote the prefill processing time. Then we have:
\begin{gather}
    \label{eq:ttft-1}
    \text{TTFT}_i = q_1^{(i)} + p_1^{(i)} = \max\{e_{i-1} - a_i, 0\} + p_1^{(i)} \\
    \label{eq:ttft-2}
    e_i = a_i + \text{TTFT}_i
\end{gather}
Specifically, $\text{TTFT}_1 = p_1$. We summarize three key characteristics of TTFT:
\begin{itemize}
    \item \textbf{Strong Predictability}: Equations~\ref{eq:ttft-1} and~\ref{eq:ttft-2} indicate that the TTFT of the $i$-th request can be uniquely determined by the arrival times and prefill processing times of the first to $i$-th requests. Since the relationship between the computation time of a prefill request and input length can be determined through profiling and interpolation~\cite{zhongDistServeDisaggregatingPrefill2024, qinMooncakeTradingMore2025, duEcoServeEnablingCosteffective2025}, the TTFT of each request can be accurately predicted.
    \item \textbf{Monotonicity}: Starting from the moment a user issues a request, the TTFT of the request can only increase monotonically with the processing time. If the current queuing delay and computation time exceed the TTFT SLO, the request can no longer meet the SLO requirement.
    \item \textbf{Sensitivity to Burst Traffic}: Consider the case where $n$ requests arrive as a burst, meaning their arrival times $a_i$ fall within a narrow interval, In this case, $a_i$ can be approximated as a constant, and the monotonic increase of $e_i$ causes the TTFT of requests 1 through $n$ to exhibit an increasing trend.
\end{itemize}

\parabf{Insight 1:} The strong predictability of TTFT enables the serving system to leverage queue information from prefill instances to accurately predict the TTFT of new requests, thereby anticipating potential violations of TTFT SLO.

\parabf{Insight 2:} The monotonicity of TTFT and its sensitivity to burst traffic imply that the serving system cannot rely on monitoring the TTFT metrics of completed requests to make instance scheduling decisions. Otherwise, this approach may lead to TTFT SLO violations for later-queued requests in bursty traffic scenarios, with no remedial actions available to bring these requests back into SLO compliance.

\subsection{TPOT}
\label{subsec:analyais-tpot}
The TPOT (Time-per-Output-Token) metric is a key indicator of the processing capability of decode instances. It represents the average waiting time between every two consecutive tokens received by the user. Let $t_{j+1}$ denote the time interval between the output tokens $o_j$ and $o_{j+1}$ of request $r$. Then, TPOT can be expressed as:
\begin{equation}
    \label{eq:tpot-1}
    \text{TPOT} = \frac{\sum_{j=2}^{m} t_j}{m-1} = \frac{\text{Decode Phase Time}}{m-1}\ (m \geq 2)
\end{equation}
Here, $t_j = p_j$ if $j > 2$. Specifically, $t_2 = q_2 + c + q_3 + p_2$. We focus our analysis on the four components of $t_2$:
\begin{itemize}
    \item KV Cache Transmission Queueing Delay $q_2$ and Decode Queueing Delay $q_3$: Having sufficient GPU memory is the prerequisite for a decode instance to fetch the cache of a decode request and execute it. However, the iterative process of LLMs makes it difficult to predict the output length of each request in advance, further complicating the estimation of the available GPU memory of decode instances at any given moment. Consequently, both queuing delays are highly unpredictable when the decode instance is under high load with limited available memory.
    \item Cache Transmission Time $c$: It can be determined by the size of the KV Cache to be transmitted and the available bandwidth.
    \item Decode Iteration Time $p_2$: The relationship between processing time and the number of tokens in the batch can be determined through profiling~\cite{zhongDistServeDisaggregatingPrefill2024}.
\end{itemize}

We summarize two key characteristics of TPOT:
\begin{itemize}
    \item \textbf{Weak Predictability}: The uncertainty in request output length makes several delays difficult to predict, resulting in the weak predictability of TPOT.
    \item \textbf{Non-monotonicity}: Equation~\ref{eq:tpot-1} shows that TPOT is determined by the generation intervals of all tokens. Thus, the TPOT of a request does not exhibit a definite monotonic relationship with processing time.
\end{itemize}

\parabf{Insight 3.} The weak predictability of TPOT makes it challenging for the serving system to accurately predict new requests' TPOT. Real-time monitoring of token generation intervals is required to detect TPOT SLO violations.

\parabf{Insight 4.} The non-monotonicity of TPOT allows decode instances to tolerate temporary workload spikes. Longer generation delays for some tokens do not always result in TPOT SLO violations.

\begin{figure}[bt]
    \centering
    \includegraphics[width=\linewidth]{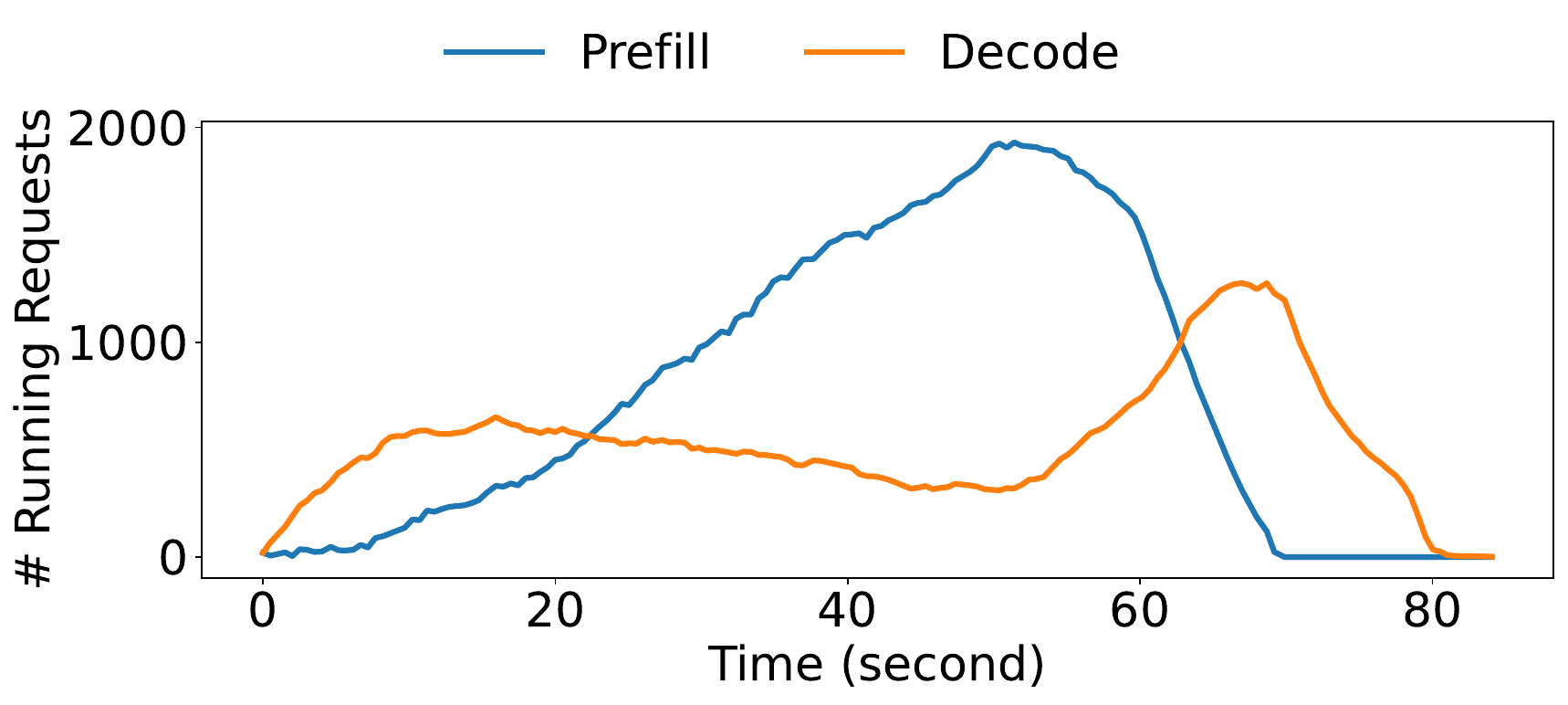}
    \caption{Prefill and Decode load over time when serving the Azure Conversation workload.}
    \label{fig:analysis-load}
\end{figure}

\subsection{Load Difference}
\label{subsec:analysis-load}
We take a clip of the Azure Conversation trace from the 20th to the 40th minute as an example to analyze the load differences between prefill and decode instances under gradually increasing input load. This clip is characterized by a rising trend in request count per minute. Figure~\ref{fig:analysis-load} illustrates the number of requests being processed by prefill and decode instances over time. Under gradually increasing workload, since requests must be processed sequentially through prefill instance followed by decode instance, the prefill instances experience an earlier onset of load increase, peak load timing, and load decline compared to decode instances.

\parabf{Insight 5.} The mandatory Prefill $\rightarrow$ Decode computation order creates temporal misalignment in peak load patterns between prefill and decode instances, offering optimization opportunities for instance scheduling under bursty traffic: When prefill load increases, some decode instances with still-low load can be temporarily scheduled for prefill computation, until decode load begins to rise, at which point more instances should be reallocated to decode computation.

\begin{figure*}[t]
    \centering
    \includegraphics[width=\linewidth]{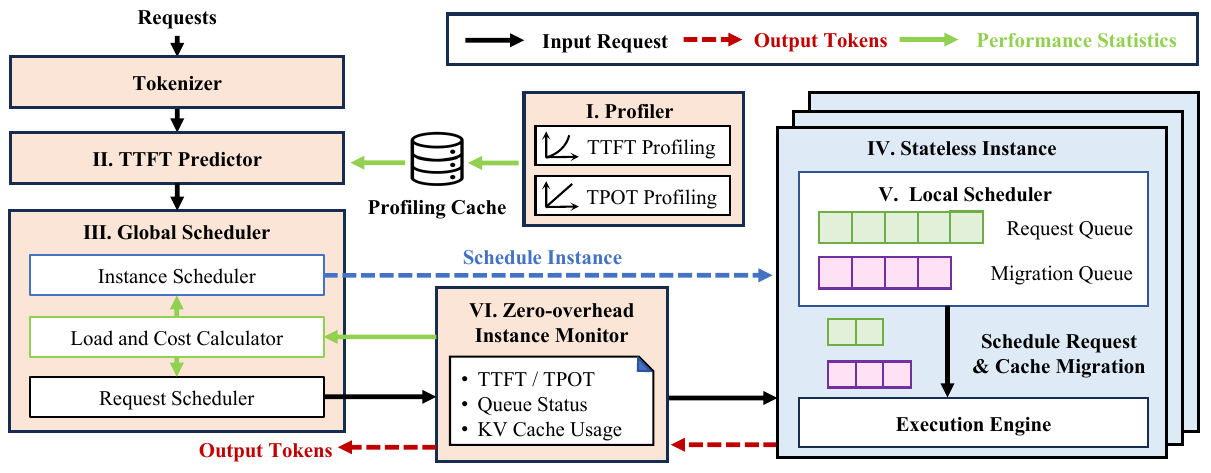}
    \caption{\sysname{} architecture overview.}
    \label{fig:design-overview}
\end{figure*}

\section{Design}
\label{sec:design}

\subsection{Overview}
\label{subsec:design-overview}

Based on the analyses in Sections~\ref{sec:motivation} and~\ref{sec:analysis}, we design \sysname{}, an adaptive scheduling engine for Prefill-Decode disaggregated architecture. Figure~\ref{fig:design-overview} illustrates the architecture of \sysname{}. \sysname{} adopts a stateless design for instances (\rom{4}), which can process both prefill and decode requests. When the cluster is initially started, the profiler (\rom{1}) performs TTFT and TPOT profiling for each instance to model their prefill and decode processing capabilities. The instance monitor (\rom{6}) collects real-time performance data such as TTFT and TPOT by recording the input and output information of each instance. When a new request arrives, the global scheduler (\rom{3}) computes the cost of dispatching the request to each instance based on the predicted TTFT from the TTFT predictor (\rom{2}) and the performance data recorded by the instance monitor, then assigns the request to the instance with the minimum cost. The local scheduler (\rom{5}) on the instance schedules request computations and KV Cache transfers in each iteration. The global scheduler also dynamically adjusts the role labels of instances based on real-time performance data, enabling rapid adaptation to load variations.

\subsection{Instance Management}
\label{subsec:design-instance-management}
\parabf{Stateless Instance.} \sysname{} designs instances to be stateless, allowing each instance to handle both prefill and decode requests. When a new request arrives, the global scheduler selects Instance A to process the prefill phase. After Instance A completes the prefill computation, the global scheduler then selects another instance, Instance B, to execute the decode stage computation. Upon receiving the decode request, Instance B pulls the KV Cache from Instance A and begins the iterative decode phase. This provides the global scheduler with greater flexibility in request and instance scheduling: (1) Each request is split into prefill and decode sub-requests, which can be scheduled independently. The global scheduler can use different scheduling strategies for these two types of requests, and may even assign both phases to the same instance if desired. (2) Prefill and decode are no longer treated as attributes of instances, but solely as attributes of requests, completely eliminating flip waiting time and instance restart time during instance scheduling.

\parabf{Instance Monitor.} Traditional performance monitoring typically requires instances to actively record and report metrics. However, query intervals can cause the performance data obtained by the scheduler during decision-making to be non-real-time, while recording performance data on inference instances also introduces additional overhead. To address this, we deploy a zero-overhead instance monitor that computes metrics such as token generation latency, queue status, and KV Cache usage on an independent component by listening to the request information dispatched by the scheduler to instances and each output token from the instances. This enables real-time performance monitoring without impacting the performance of the inference instances.

\parabf{Processing Capability Modeling.} To further decouple the global scheduler from the computing instances, enabling the scheduler to remain agnostic to low-level deployment details such as the specific parallelization strategies used by instances, we profile the prefill and decode processing capabilities of each instance when the cluster is first initialized. The profiler will send requests of varying lengths to the instances, records the TTFT, and models the prefill computation capability of each instance by fitting a quadratic curve to the relationship between input length and TTFT. Then, by sending requests with extremely long output length requirements, it records how the token generation interval varies with the number of tokens in the batch to model the instance's decode capability. This information is cached to disk and can be reused in subsequent cluster startups. If an instance's computational capability changes - for example, due to a change in deployment configuration - only that specific instance needs to be re-profiled.

\subsection{SLO-aware Global Request Scheduling}
\label{subsec:design-global-request-scheduling}
Building on the minimum-load scheduling strategy, \sysname{} further designs its request scheduling strategy to be SLO-aware, meaning that the scheduler considers real-time TTFT and TPOT of existing requests and makes decisions in conjunction with the SLO targets. As shown in Algorithm~\ref{alg:global-request-scheduling}, when a new request arrives, if it is a decode request and the instance that processed the prefill stage of this request has been reassigned to decode instance, the request is directly dispatched to that instance to avoid KV cache transfer. Otherwise, the scheduler searches for the lowest-cost instance that can also meet the SLO requirements. If such an instance does not exist, the scheduler will try to flip an instance based on cluster load. If the request still cannot be satisfied, it falls back to the instance with the minimum cost.

\begin{algorithm}[t]
    \caption{SLO-aware Global Request Scheduling}
    \label{alg:global-request-scheduling}
\begin{algorithmic}[1]
    \STATE {\bfseries Input:} Request $r$, Instances $I$
    \STATE {\bfseries Output:} Target instance $t$

    \COMMENT{Prefill instance has been flipped to decode}
    \IF{$r.type = \text{Decode} = r.prefill\_instance.role$}
        \RETURN $r.prefill\_instance$
    \ENDIF
    
    \COMMENT{1. Compute cost for each instance}
    \FORALL{$i \in I$}
        \STATE $costs[i]$ $\gets$ GetCost($r$, $i$)
    \ENDFOR

    \COMMENT{2. Find minimal-cost instance satisfying SLO}
    \STATE $(best\_i, best\_cost) \gets \min\limits_{(i,c) \in costs,~SLO(i,c)=\text{True}} (i,c)$

    \IF{$best\_i \neq None$}
        \RETURN $best\_i$
    \ENDIF

    \COMMENT{3. Try to flip an instance based on load condition}
    \IF{$r.type=\text{Decode}$ \OR Decode load is low}
        \IF{$(new\_i \gets \text{Flip}(r.type)) \neq None$}
            \RETURN $new\_i$
        \ENDIF
    \ENDIF
    
    \COMMENT{4. Fallback: select instance with minimal cost}
    \RETURN ${\operatorname{arg\,min}}_{i \in I} \ costs[i]$
\end{algorithmic}
\end{algorithm}

The computational cost for request $r$ on instance $i$ is calculated as follows, where $P$ and $D$ are the prefill and decode requests on instance $i$, and $L(r)$ is the length of request $r$:
\begin{itemize}
    \item For a prefill request $r$, the computational cost is defined as a tuple $$\left( \sum_{r_d \in D} L(r_d),\ \sum_{r_p \in P \cup \{r\}} T(r_p, i) \right)$$
    
    Here, $T(r, i)$ represents the prefill processing time of request $r$ on instance $i$, provided by the TTFT predictor based on the instance's TTFT profiling data. The first component of the cost represents the total number of decode tokens currently being processed on the instance. This encourages the scheduler to favor instances that are only handling prefill requests over others, avoiding dispatching new prefill requests to instances still processing remaining decode requests whenever possible to avoid interference. The second component indicates that instances with smaller prefill processing times have lower costs.
    \item For a decode request $r$, the cost is similarly defined as $$\left( \sum_{r_p \in P} L(r_p),\ \sum_{r_d \in D \cup \{r\}} L(r_d) - \operatorname{MT}(i, \text{SLO}_\text{TPOT}) \right)$$
    
    Here, $\operatorname{MT}(i, \text{SLO}_\text{TPOT})$ denotes the maximum number of tokens instance $i$ can compute concurrently under the given TPOT SLO, derived from the instance's TPOT profiling data. The design of the first component is similar to that for prefill costs: the scheduler tries to avoid dispatching decode requests to instances that are still processing remaining prefill requests. The second component represents the distance between the current number of tokens being processed on the instance and its maximum capacity; a smaller value indicates a lighter load.
\end{itemize}

For the SLO check function, it simply checks whether the second component of the prefill cost exceeds the TTFT SLO threshold, or whether the second component of the decode cost is greater than 0. The calculation of instance load and flipping operations will be introduced in Section~\ref{subsec:design-instance-scheduling}.

\subsection{Local Request Scheduling}
\label{subsec:design-local-request-scheduling}
When a new request arrives, the local scheduler first checks whether KV Cache migration is required. If so, the request is placed in the migration queue and moved to the request queue after migration completes. The local scheduler adopts a FCFS policy for KV Cache migration, and uses the chunked prefill scheduling strategy~\cite{agrawalTamingThroughputLatencyTradeoff2024} for requests: Under a given batch size, decode requests are prioritized to be included in the running batch. If there is remaining space, chunked prefill requests are added. This strategy enables instances to begin processing new types of requests as soon as possible during role flipping, avoiding the situation where requests queued before instance flipping block the execution of new requests after flipping.

\begin{algorithm}[t]
\caption{Global Scheduler Monitoring Loop}
\label{alg:monitor-loop}
\begin{algorithmic}[1]
\STATE {\bfseries Input:} Prefill instances $I_P$, Decode instances $I_D$
\FOR{every update interval}
    \STATE $L_P, L_D \gets \text{GetLoad}(I_P), \text{GetLoad}(I_D)$
    \IF{$L_D \geq L_\text{EXPAND}$ \OR $L_P \leq L_\text{SHRINK} \leq L_D$}
        \STATE Flip(Decode)
    \ENDIF
\ENDFOR
\end{algorithmic}
\end{algorithm}

\subsection{SLO-aware Instance Scheduling}
\label{subsec:design-instance-scheduling}
\parabf{Flipping Timing.} The instance scheduling strategy employed by \sysname{} is also SLO-aware. Algorithm~\ref{alg:monitor-loop} describes the scheduler's monitoring loop. Here, the prefill load of an instance is defined as the ratio of total estimated prefill processing time to TTFT SLO, while the decode load is defined as the ratio of the average latency of tokens generated between the update interval to TPOT SLO. The load of the instance pool is the average load of all instances within it.
\begin{itemize}
    \item Instance scheduling from decode to prefill occurs during the prefill request scheduling process (line 10 of Algorithm~\ref{alg:global-request-scheduling}): Based on Insights 1 and 2, when the scheduler predicts that the current prefill instances cannot meet the TTFT SLO requirement for a new request, it will attempt to reassign decode instances to the prefill instance pool.
    \item Instance scheduling from prefill to decode occurs in the following situations: (1) During the decode request scheduling process (line 10 of Algorithm~\ref{alg:global-request-scheduling}); (2) When the scheduler detects that the average load of decode instances exceeds a threshold over a period of time (line 4 of Algorithm~\ref{alg:monitor-loop}, $L_D \geq L_\text{EXPAND}$); (3) When prefill instances are under low load while decode instances are not idle, idle prefill instances are added to decode computation to free up computing resources as quickly as possible in anticipation of potential future bursty traffic. (line 4 of Algorithm~\ref{alg:monitor-loop}, $L_P \leq L_\text{SHRINK} \leq L_D$).
\end{itemize}

\begin{algorithm}[t]
\caption{Instance Scheduling}
\label{alg:instance-scheduling}
\begin{algorithmic}[1]
\STATE {\bfseries Input:} Source instances $S$, target instances $T$, direction flag $d \in \{\text{P2D}, \text{D2P}\}$
\STATE {\bfseries Output:} Flipped instance $t$ or \texttt{None}

\IF{$d = \text{P2D}$ \AND $t_{\text{now}} - t_{\text{last\_flip}} < \text{COOLDOWN}$}
    \RETURN None
\ENDIF

\IF{$|S| > 1$}
    \FORALL{$instance \in S$}
        \STATE $costs[instance] \gets \text{GetFlipCost}(instance)$
    \ENDFOR
    \STATE $t \gets {\operatorname{arg\,min}}_{i \in S} \ costs[i]$
    \STATE $S,\ T \gets S - \{t\},\ T \cup \{t\}$
    \RETURN $t$
\ENDIF

\RETURN None
\end{algorithmic}
\end{algorithm}

\parabf{Flipping Target.} Algorithm~\ref{alg:instance-scheduling} details the instance scheduling process. The scheduler flips the instance with the minimum flipping cost. To prevent oscillation in instance assignment, we introduce a cooldown mechanism to avoid overly frequent adjustments. Based on the analysis in Section~\ref{sec:analysis}, the cooldown mechanism is only applied to P$\rightarrow$D process, since the load of decode instances requires a period of observation to determine, and the weak predictability of TPOT means that a slight lag in P$\rightarrow$D scheduling is tolerable. In contrast, TTFT, due to its strong predictability and sensitivity to traffic spikes, requires rapid instance scheduling.

The flipping cost for a prefill instance is defined as $$\left( I[D=\emptyset],\ \sum_{r_p \in P} T(r_p, i) \right)$$

Similarly, the flipping cost for a decode instance is $$\left( I[P=\emptyset],\ \sum_{r_d \in D} L(r_d) \right)$$

Here, $I$ is the indicator function. The first component is used to check whether requests of the other type still exist on current instance. If they do, it indicates that the instance's role has been flipped previously and the flipping is not yet complete. The scheduler prioritizes flipping instances of this type, effectively revoking the previous flipping operation. The second component indicates that instances with a lighter load have a lower cost.

\parabf{Scheduling in Overload Scenario.}
Scenarios in which both prefill and decode tasks are overloaded are not the primary optimization target for \sysname{}. However, to prevent instance scheduling oscillations in such scenarios, \sysname{} prioritizes allocating compute resources to the decode requests. Specifically, in Algorithm~\ref{alg:global-request-scheduling}, before a D$\rightarrow$P flip, the scheduler checks the load of the decode instance and aborts the flip if the decode load is high, whereas P$\rightarrow$D flips proceed without prefill load checks. The core rationale for this design is to avoid scenarios where a large number of requests occupy memory resources without progressing beyond the prefill phase. Existing work~\cite{qinMooncakeKVCachecentricDisaggregated2024} has also proposed request scheduling schemes for overloaded scenarios, but the design of such schemes is beyond the scope of this paper.

\subsection{Implementation Details}
\sysname{} is currently built upon vLLM~\cite{kwonEfficientMemoryManagement2023} and utilizes NIXL~\cite{nvidiaAidynamoNixl2025} for KV Cache transmission. Components such as the Profiler and Monitor are transparent to the backend instances, enabling the design of \sysname{} to be extended to other inference engines. The only requirement is that the inference engine is implemented as stateless and capable of KV Cache transmission with any arbitrary instance. Alternatively, distributed KV Cache storage solutions like Mooncake~\cite{qinMooncakeKVCachecentricDisaggregated2024} can be used to further optimize KV Cache transmission and global cache reuse.
\begin{figure*}[t]
    \centering
    \begin{subfigure}
        \centering
        \includegraphics[width=\linewidth]{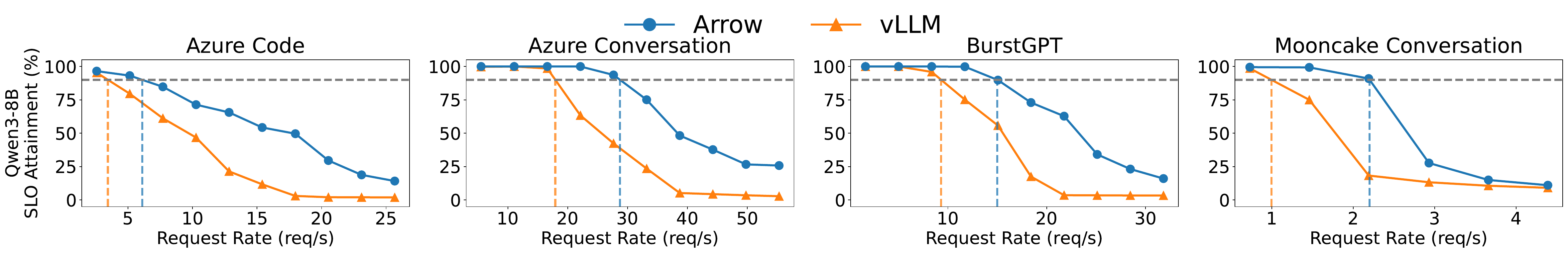}
    \end{subfigure}
    \begin{subfigure}
        \centering
        \includegraphics[width=\linewidth]{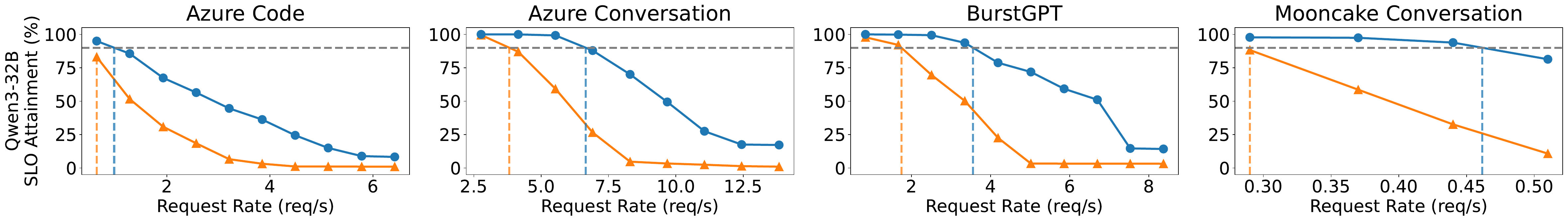}
    \end{subfigure}
    \begin{subfigure}
        \centering
        \includegraphics[width=0.75\linewidth]{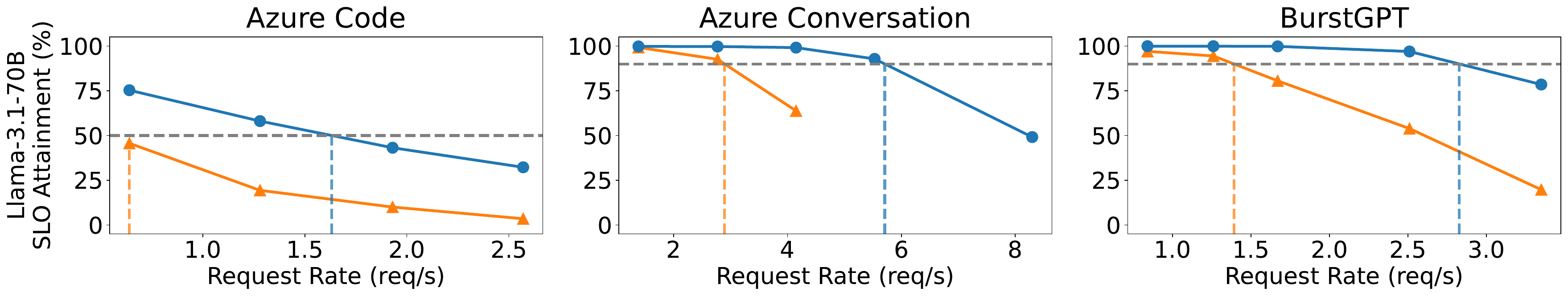}
    \end{subfigure}
    \caption{Performance of different LLM serving systems employing different models under various traces and request rates.}
    \label{fig:eval-performance}
\end{figure*}

\section{Evaluation}
\label{sec:evaluation}
In this section, we evaluate the performance of \sysname{} with state-of-the-art PD-disaggregated systems on real-world workloads and show the effectiveness of its components.

\subsection{Experimental Setup}
\label{subsec:evaluation-setup}
\parabf{Testbed.} We evaluate \sysname{} on two servers, each equipped with 8 NVIDIA H20 141GB GPUs, 2$\times$200Gbps InfiniBand NICs, 96-core CPUs, and 2048GB of host memory.

\parabf{Model.} We evaluated the performance of \sysname{} on models of varying sizes, including Qwen3-8B, Qwen3-32B~\cite{yangQwen3TechnicalReport2025}, and Llama-3.1-70B~\cite{grattafioriLlama3Herd2024}.

\parabf{Workloads.} We choose the four LLM serving traces introduced in Section~\ref{sec:introduction} as test workloads. Detailed descriptions of these traces and SLO settings are included in Appendix~\ref{app:workload}.

\parabf{Baseline.} We use vLLM v0.11~\cite{kwonEfficientMemoryManagement2023} as baseline system, as it represents state-of-the-art Prefill-Decode disaggregated inference serving system. We launched 4 prefill instances and 4 decode instances to handle the requests.

\parabf{Metrics. } We use \textit{SLO attainment} as the major metric. Under a specific SLO setting, we are concerned with the maximum request rate the system can handle. We set the SLO attainment target to 90\%, which is a common setting in previous work~\cite{patelSplitwiseEfficientGenerative2024,zhongDistServeDisaggregatingPrefill2024}.

\parabf{Evaluation Workflow.} We adopt the same evaluation workflow as previous works~\cite{wangBurstGPTRealWorldWorkload2025, qinMooncakeTradingMore2025}, assessing the performance of different serving systems by replaying service traces. To evaluate system performance under different request rates, we multiply the timestamps by a constant to simulate varying request rates.

\subsection{End-to-End Performance}
We compared the performance of \sysname{} and the baseline system across four real-world serving traces. Figure~\ref{fig:eval-performance} shows the test results. On the Qwen3-8B model, \sysname{} achieves $1.60\times$\textasciitilde$2.21\times$ higher sustainable request rates compared to vLLM. This is because \sysname{} leverages its SLO-aware request and instance scheduling strategy to effectively balance the computational demands of prefill and decode tasks, striving to meet both TTFT and TPOT SLOs for requests simultaneously. Similar results can be observed in tests with the larger model Qwen3-32B, where \sysname{} improves request goodput by $1.59\times$\textasciitilde$2.03\times$ compared to vLLM. vLLM consistently failed to reach 90\% SLO attainment on the Azure Code dataset, which has significant bursty traffic, and the Mooncake Conversation dataset, which contains ultra-long request lengths. In contrast, \sysname{} utilizes the strong predictability of TTFT to promptly allocate more instances to prefill computation when bursty traffic arrives, preventing a large number of burst requests from violating TTFT SLO due to long queuing delays. For the Llama-3.1-70B model, we set the tensor parallelism size to 2. Due to the high load of the Azure Code dataset, we set its SLO attainment target to 50\%. On the Azure Conversation dataset, vLLM experienced KV Cache transfer failures under high load, preventing completion of the test. The Mooncake Conversation dataset was not tested on Llama-3.1-70B due to its ultra-long input sequences, which easily exceed the memory capacity of the serving systems. In the 70B model tests, \sysname{} yields $1.97\times$\textasciitilde$2.55\times$ improvements over the baseline, effectively increasing the system's serving capacity.

\subsection{Ablation Study}
\begin{figure}[t]
  \centering
  \includegraphics[width=\linewidth]{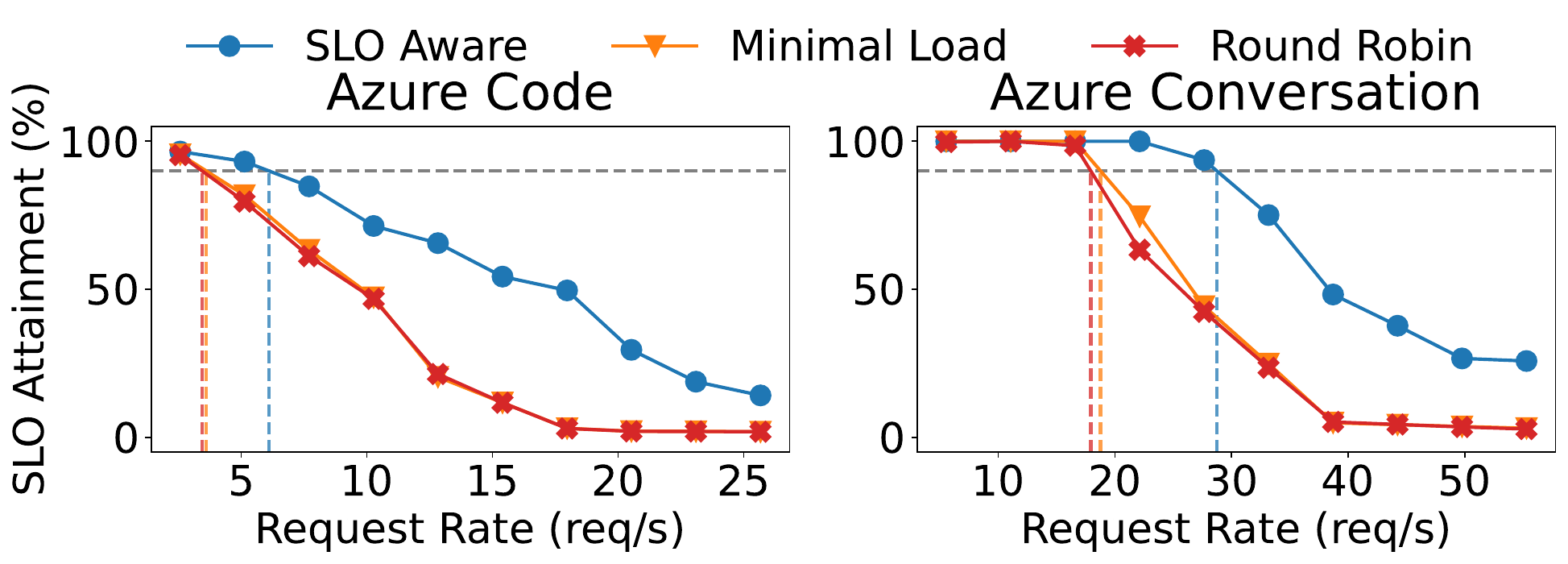}
  \caption{Performance of different scheduling strategies.}
  \label{fig:eval-strategy}
\end{figure}

In this section, we study the effectiveness of \sysname{}'s adaptive scheduling strategy. We compare the performance of three scheduling strategies on the Qwen3-8B model: (1) SLO Aware, which is the strategy used by \sysname{} and includes both request scheduling strategy and instance scheduling strategy from Section~\ref{subsec:design-global-request-scheduling} and~\ref{subsec:design-instance-scheduling}; (2) Minimal Load, which only includes the minimum-load request scheduling strategy; and (3) Round Robin. The results are shown in Figure~\ref{fig:eval-strategy}.

On the Azure Code dataset, the SLO Aware strategy used by \sysname{} achieves $1.69\times$ higher request serving rate compared to the Minimal Load strategy, demonstrating the effectiveness of adaptive instance scheduling. Compared to the Round Robin strategy, the Minimal Load request scheduling strategy achieves up to a 2.2\% improvement in SLO attainment. For the Azure Conversation dataset, the SLO Aware strategy achieves $1.53\times$ higher request serving rate than the Minimal Load strategy, serving 10 additional requests per second. The Minimal Load strategy achieves up to 11.2\% improvement in SLO attainment compared to Round Robin strategy, proving that minimal-load scheduling can more closely approximate the optimal scheduling strategy.

\subsection{Scalability}
\begin{figure}[t]
  \centering
  \includegraphics[width=\linewidth]{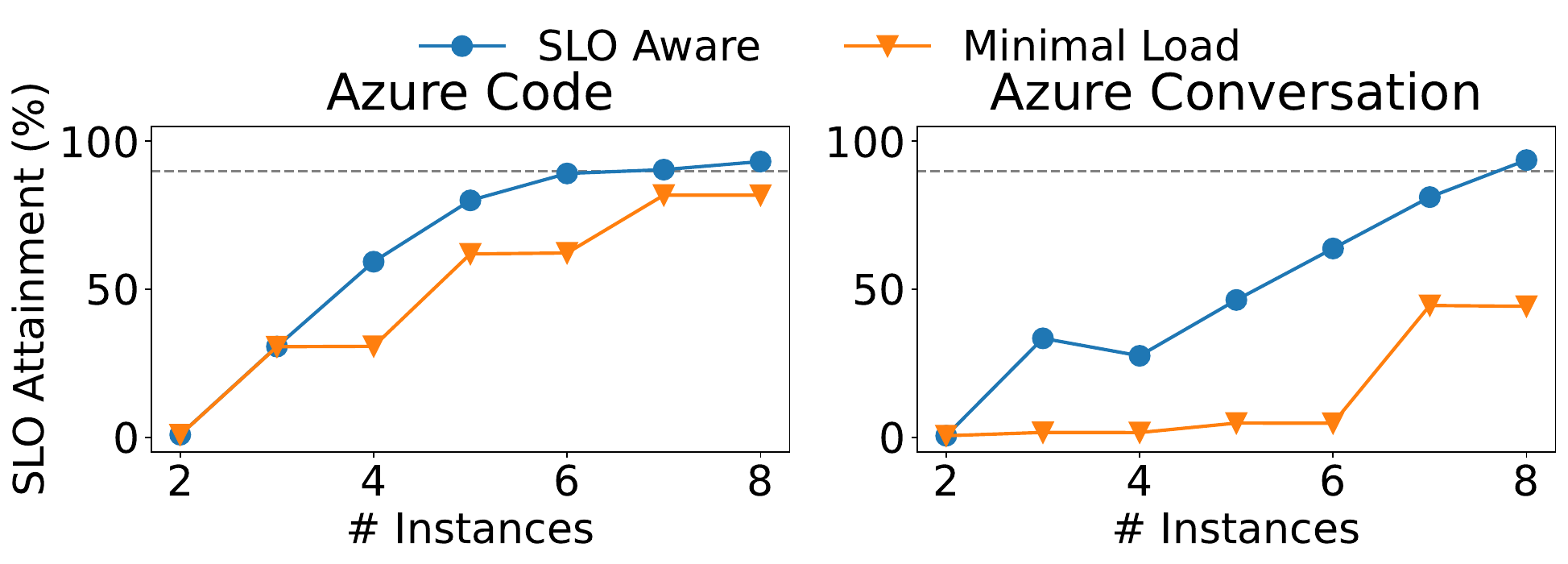}
  \caption{Performance under different number of instances.}
  \label{fig:eval-scalability-count}
\end{figure}

\parabf{Strategy Scalability.} We compare the SLO attainment of the SLO Aware and Minimal Load scheduling strategies under varying instance counts to evaluate the scalability of \sysname{}'s scheduling strategy. The results are shown in Figure~\ref{fig:eval-scalability-count}. By employing a flexible instance scheduling strategy, \sysname{} can fully utilize computational resources to meet the demands of both prefill and decode phases, enabling the serving system to achieve significant improvements in SLO attainment as the number of instances increases. In contrast, traditional static PD ratio configurations are prone to hitting either prefill or decode computation bottlenecks under resource constraints, making it difficult to satisfy both TTFT and TPOT SLOs simultaneously. For example, in tests on the Azure Conversation dataset, when the number of instances increased from 2 to 6, the number of both prefill and decode instances in the cluster using the Minimal Load strategy increased by 2, but the SLO attainment rate improved only marginally. In contrast, the SLO Aware scheduling strategy effectively adjusts the number of instances for both types, achieving an SLO attainment rate exceeding 60\%. The experimental results demonstrate that our adaptive scheduling strategy exhibits strong universality and scalability, enabling efficient computational resource utilization across different hardware environments.

\parabf{Scheduler Scalability.} To test whether the centralized Tokenizer and Global Scheduler could become system bottlenecks, we measured the tokenization latency and scheduling latency under different request rates on the Azure Conversation dataset. The results are shown in Figure~\ref{fig:eval-scalability-scheduler-heterogeneous}(Left). As the request rate gradually increases, the tokenization latency shows a slight rise, while the scheduling latency remains almost unchanged. Compared to the second-level processing time of the prefill stage, the increase in latency for tokenization and scheduling is negligible. In scenarios involving long contexts or systems with extremely high concurrency, if tokenization becomes a bottleneck, it can be addressed by deploying multiple tokenization service nodes. The global scheduler, when dispatching requests, only needs to perform simple calculations for the scheduling cost, thus it is unlikely to become a performance bottleneck.

\subsection{Heterogeneous Deployment}
\begin{figure}[b]
    \centering
    \begin{minipage}[b]{0.49\columnwidth}
        \centering
        \includegraphics[width=\columnwidth]{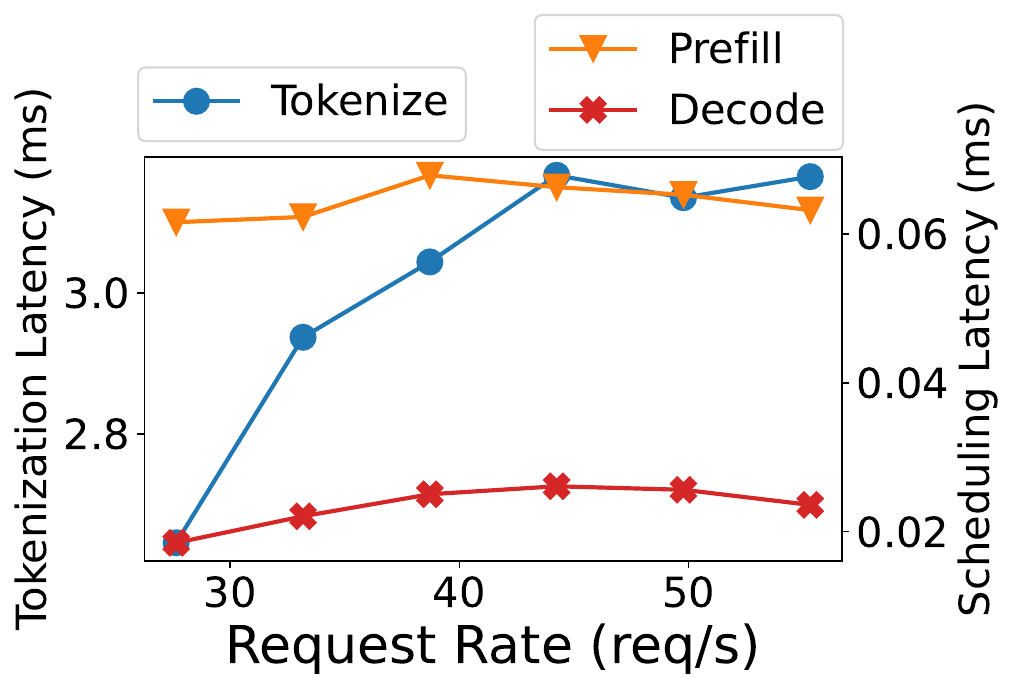}
    \end{minipage}
    \hfill
    \begin{minipage}[b]{0.49\columnwidth}
        \centering
        \includegraphics[width=\columnwidth]{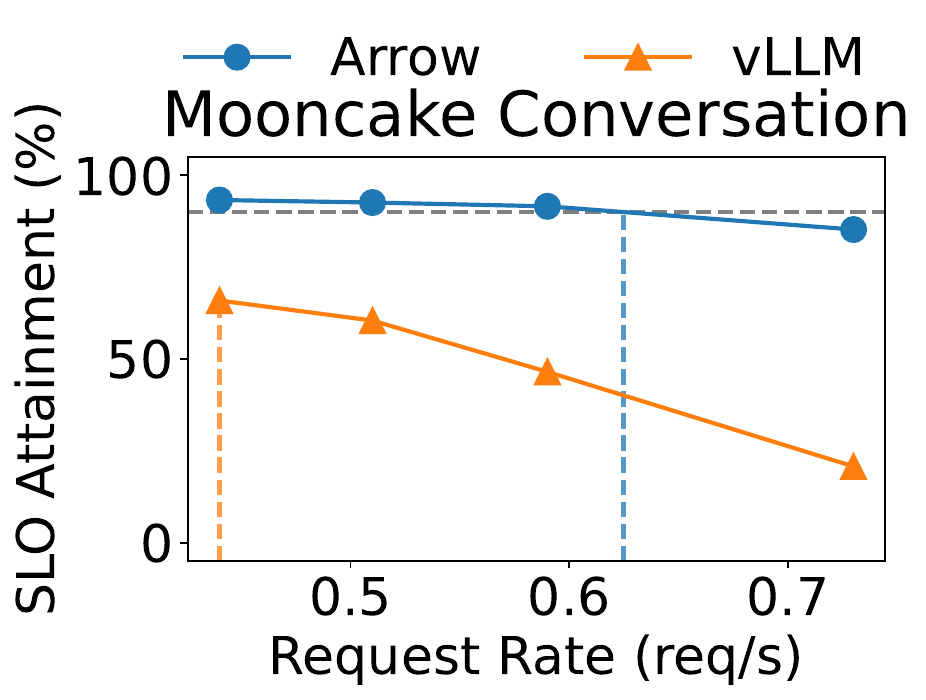}
    \end{minipage}
    \caption{(Left) Scheduling latency under different request rates. (Right) Performance results when prefill and decode instances use different parallelization strategies.}
    \label{fig:eval-scalability-scheduler-heterogeneous}
\end{figure}

\sysname{}'s scheduler directly assigns prefill or decode instances to handle requests of the other type, which may cause these instances to operate under suboptimal parallelization strategies. The rationale behind this design is to enable rapid response to traffic spikes without waiting for lengthy instance restarts that could take minutes. When the cluster load decreases, these instances are returned to their original pools. We conducted a heterogeneous instance deployment test on the Qwen3-32B model, comparing the performance of \sysname{}'s scheduling strategy with traditional static Prefill/Decode ratio configuration schemes. We set tensor parallelism size to 1 for prefill instances and 2 for decode instances. The results are shown in Figure~\ref{fig:eval-scalability-scheduler-heterogeneous}(Right). It can be observed that although \sysname{}'s immediate instance scheduling does not operate instances under their optimal parallelization strategies, it still effectively enhances the system's serving capacity. Existing work~\cite{chenGygesDynamicCrossInstance2025} has proposed several dynamic parallelization strategy switching schemes, which could be integrated into \sysname{} to improve serving performance in heterogeneous Prefill-Decode environments. We leave this integration as future work.

\section{Related Work}

\parabf{LLM Serving.} Existing works have optimized LLM serving systems from multiple perspectives, including kernel~\cite{daoFLASHATTENTIONFastMemoryefficient2022, kaoFLATOptimizedDataflow2023}, KV Cache management~\cite{geModelTellsYou2023, kwonEfficientMemoryManagement2023, liSnapKVLLMKnows2024}, and batching strategy~\cite{agrawalTamingThroughputLatencyTradeoff2024, yuOrcaDistributedServing2022}. Among these, Orca~\cite{yuOrcaDistributedServing2022} employs an iteration-level scheduling strategy to reduce request queuing latency, and Sarathi-Serve~\cite{agrawalTamingThroughputLatencyTradeoff2024} implements chunked prefill to improve compute utilization during the decode phase. These works are orthogonal to our work and have already been integrated into \sysname{}. To avoid interference between the prefill and decode phases, ShuffleInfer~\cite{huShuffleInferDisaggregateLLM2025}, Splitwise~\cite{patelSplitwiseEfficientGenerative2024}, and DistServe~\cite{zhongDistServeDisaggregatingPrefill2024} proposed the Prefill-Decode disaggregated inference architecture. EPD disaggregation~\cite{singhEfficientlyServingLarge2025} further extends this architecture to multi-modal models. However, their static PD ratio configurations are prone to SLO violations when handling varying workloads. In contrast, \sysname{} proposes an innovative SLO-aware scheduling strategy that can effectively improve serving capacity while meeting the given SLO settings.

\parabf{PD-disaggregation Optimization.} As the effectiveness of the PD-disaggregated architecture has been widely validated, numerous optimization efforts for the PD-disaggregated architecture have recently emerged. Mooncake~\cite{qinMooncakeTradingMore2025} and MemServe~\cite{huMemServeContextCaching2024} deploy a distributed KV Cache pool to enable cache reuse. D\'ej\`aVu~\cite{stratiDejaVuKVcacheStreaming2024} implements a set of high-performance KV Cache streaming APIs to reduce the KV Cache transmission overhead. These solutions can be integrated into \sysname{} to further improve its performance. Other works have optimized the PD-disaggregated architecture from perspectives including parallelization strategies~\cite{wuLoongServeEfficientlyServing2024,zhongDistServeDisaggregatingPrefill2024}, resource utilization~\cite{liangInjectingAdrenalineLLM2025, ruanDynaServeUnifiedElastic2025, hongSemiPDEfficientLLM2025}, and deployment costs~\cite{duEcoServeEnablingCosteffective2025}. Our work proposes an effective request and instance scheduling strategy that leverages TTFT's strong predictability and TPOT's non-monotonicity, thereby addressing workload diversity.

\parabf{Request Scheduling.} Existing works have optimized request scheduling in LLM serving systems for various objectives, including throughput~\cite{yuOrcaDistributedServing2022, chengEnablingEfficientBatch2024, wuFastDistributedInference2024}, load balancing~\cite{srivatsaPrebleEfficientDistributed2024, sunLlumnixDynamicScheduling2024}, and fairness~\cite{shengFairnessServingLarge2024}. Recent studies have also proposed diverse request scheduling optimizations for PD-disaggregated architecture, considering aspects such as cache~\cite{qinMooncakeTradingMore2025, huMemServeContextCaching2024}, SLO settings~\cite{duEcoServeEnablingCosteffective2025}, and instance load~\cite{zhongDistServeDisaggregatingPrefill2024, huShuffleInferDisaggregateLLM2025}. We design an SLO-aware scheduling strategy based on the minimal-load scheduling policy to enable adaptive request dispatching.

\section{Conclusion}
To tackle load fluctuations in LLM serving systems, we design \sysname{}, an efficient and adaptive scheduler that dynamically schedules requests and instances based on cluster load. \sysname{} employs stateless inference instances and SLO-aware load assessment to enable responsive instance reconfiguration, while performing adaptive request dispatching and instance scheduling based on SLO settings and real-time performance metrics. Evaluations on multiple real-world datasets demonstrate that \sysname{} can effectively improve system serving capacity compared to existing solutions.

\bibliography{reference}
\bibliographystyle{mlsys2025}

\clearpage
\appendix

\begin{table*}[!t]
    \caption{Workloads and SLO settings in evaluation.}
    \label{tab:workload}
    \centering
    \begin{tabular}{cccccc}
    \toprule
    \textbf{Trace}                      & \textbf{\# Requests}   & \textbf{Model} & \textbf{Request Rate (req/s)} & \textbf{TTFT} & \textbf{TPOT} \\ \hline
    \multirow{3}{*}{Azure Code}         & \multirow{3}{*}{8819}  & Qwen3-8B       & 2.6 - 25.7                    & 6s            & 0.1s          \\
                                        &                        & Qwen3-32B      & 0.6 - 6.4                     & 10s           & 0.125s        \\
                                        &                        & Llama-3.1-70B  & 0.6 - 2.6                     & 10s           & 0.2s          \\ \hline
    \multirow{3}{*}{Azure Conversation} & \multirow{3}{*}{19366} & Qwen3-8B       & 5.5 - 55.3                    & 3s            & 0.15s         \\
                                        &                        & Qwen3-32B      & 2.8 - 13.8                    & 6s            & 0.175s        \\
                                        &                        & Llama-3.1-70B  & 1.4 - 8.3                     & 3s            & 0.2s          \\ \hline
    \multirow{3}{*}{BurstGPT clip}      & \multirow{3}{*}{6009}  & Qwen3-8B       & 1.7 - 31.8                    & 1s            & 0.075s        \\
                                        &                        & Qwen3-32B      & 0.8 - 8.4                     & 2s            & 0.1s          \\
                                        &                        & Llama-3.1-70B  & 0.8 - 3.4                     & 2s            & 0.15s         \\ \hline
    \multirow{2}{*}{Mooncake clip}      & \multirow{2}{*}{1756}  & Qwen3-8B       & 0.7 - 4.4                     & 60s           & 0.2s          \\
                                        &                        & Qwen3-32B      & 0.3 - 0.5                     & 150s          & 0.2s          \\
    \bottomrule
    \end{tabular}
\end{table*}

\section{Evaluation Workload}
\label{app:workload}
\begin{figure}[h]
    \centering
    \includegraphics[width=\linewidth]{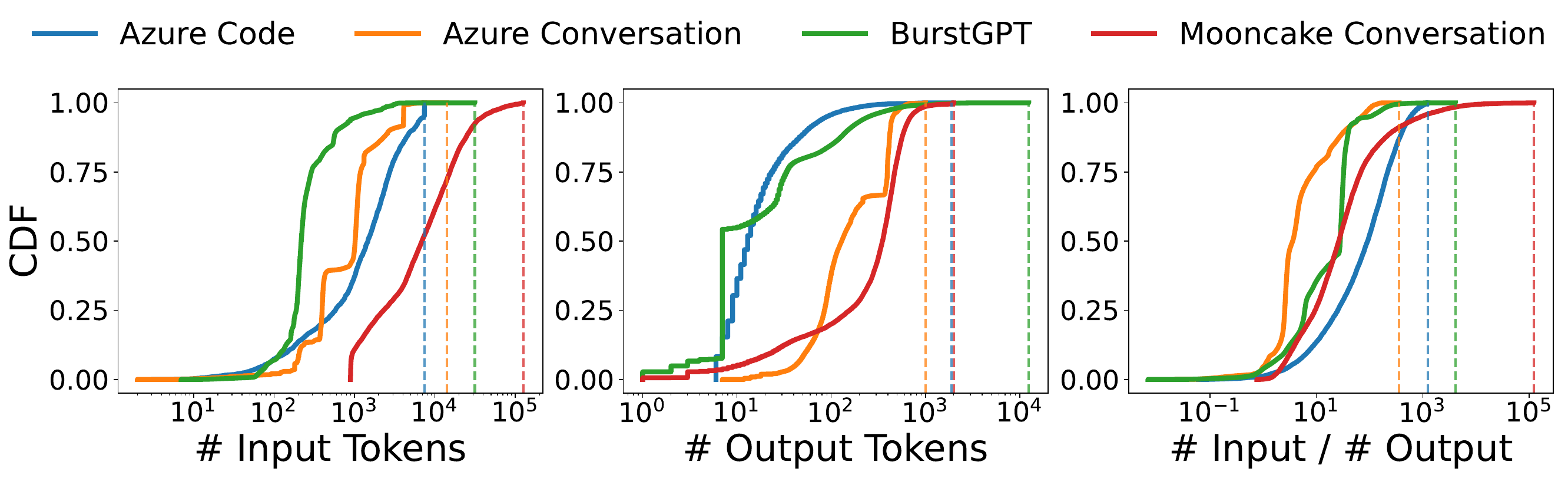}
    \caption{Input and output length distribution of different traces.}
    \label{fig:motivation-length-distribution}
\end{figure}

We choose four real-world LLM serving traces as the workload. Each trace records request information processed by the inference serving system over a period, including arrival time and input/output lengths. Figure~\ref{fig:motivation-length-distribution} presents the cumulative distribution functions (CDFs) of input and output lengths across these traces. Detailed information about the workloads used in evaluation is shown in Table \ref{tab:workload}.

\begin{itemize}
    \item \textbf{Azure LLM Inference Traces}~\cite{patelSplitwiseEfficientGenerative2024}: It is a 1-hour serving trace collected from Azure LLM inference services, including both coding and conversation scenarios.
    \item \textbf{BurstGPT}~\cite{wangBurstGPTRealWorldWorkload2025}: It is an LLM serving workload with 5.29 million traces from regional Azure OpenAI GPT services over 121 days. We take a 1-hour clip from the original trace for evaluation.
    \item \textbf{Mooncake Conversation Trace}~\cite{qinMooncakeTradingMore2025}: It is a 1-hour conversation trace containing a significant portion of long context requests. Replaying the full trace will exceed the serving capacity of all the tested systems, so we only take the first ten minutes of requests for evaluation. 
\end{itemize}

\end{document}